\documentclass[
    ,final            
  ]
  {aipproc}

\layoutstyle{6x9}

\begin{document}

\title{Towards Real-time Classification of Astronomical Transients}


\author{A. Mahabal}{
  address={California Institute of Technology, MC 105-24, 1200 E. California Bl.,
Pasadena, CA 91125, USA}
}

\author{S. G. Djorgovski}{
  address={California Institute of Technology, MC 105-24, 1200 E. California Bl.,
Pasadena, CA 91125, USA}
}

\author{R. Williams}{
  address={California Institute of Technology, MC 105-24, 1200 E. California Bl.,
Pasadena, CA 91125, USA}
}

\author{A. Drake}{
  address={California Institute of Technology, MC 105-24, 1200 E. California Bl.,
Pasadena, CA 91125, USA}
}

\author{C. Donalek}{
  address={California Institute of Technology, MC 105-24, 1200 E. California Bl.,
Pasadena, CA 91125, USA}
}

\author{M. Graham}{
  address={California Institute of Technology, MC 105-24, 1200 E. California Bl.,
Pasadena, CA 91125, USA}
}

\author{B. Moghaddam}{
  address={Jet Propulsion Laboratory, 4800 Oak Grove Dr., Pasadena, CA 91109, USA}
}

\author{M. Turmon}{
  address={Jet Propulsion Laboratory, 4800 Oak Grove Dr., Pasadena, CA 91109, USA}
}

\author{J. Jewell}{
  address={Jet Propulsion Laboratory, 4800 Oak Grove Dr., Pasadena, CA 91109, USA}
}

\author{A. Khosla}{
  address={California Institute of Technology, MC 105-24, 1200 E. California Bl.,
Pasadena, CA 91125, USA}
}

\author{B. Hensley}{
  address={California Institute of Technology, MC 105-24, 1200 E. California Bl.,
Pasadena, CA 91125, USA}
}

\begin{abstract}
Exploration of time domain is now a vibrant area of research in astronomy,
driven by the advent of digital synoptic sky surveys.  While panoramic surveys
can detect variable or transient events, typically some follow-up observations
are needed; for short-lived phenomena, a rapid response is essential.  Ability
to automatically classify and prioritize transient events for follow-up
studies becomes critical as the data rates increase.  We have been developing
such methods using the data streams from the Palomar-Quest survey, the
Catalina Sky Survey and others, using the VOEventNet framework.  The goal is
to automatically classify transient events, using the new measurements,
combined with archival data (previous and multi-wavelength measurements), and
contextual information (e.g., Galactic or ecliptic latitude, presence of a
possible host galaxy nearby, etc.); and to iterate them dynamically as the
follow-up data come in (e.g., light curves or colors).  We have been
investigating Bayesian methodologies for classification, as well as
discriminated follow-up to optimize the use of available resources, including
Naive Bayesian approach, and the non-parametric Gaussian process regression.
We will also be deploying variants of the traditional machine learning
techniques such as Neural Nets and Support Vector Machines on datasets of
reliably classified transients as they build up. 
\end{abstract}

\maketitle


\section{Introduction}

Time domain astronomy has rapidly emerged as one of the more exciting areas of
research in astronomy. It touches on a number of important scientific
directions, ranging from exploration of the Solar System to cosmology. Besides
the objects that move (e.g., asteroids, TNOs, KBOs), the types of transients
we are likely to encounter include SNe (cosmological standard candles, as well
as endpoints of stellar evolution), GRB orphan afterglows (which constrain the
beaming models), variable stars of all sorts (probes of stellar astrophysics
and Galactic structure), AGN (as a method of finding QSOs and constraining
their fueling mechanisms and lifetimes), etc.  There are classes of variable
events which are expected or suspected to occur, but for which there is only a
limited evidence in hand, e.g., tidal disruption of stars by otherwise
quiescent supermassive black holes \cite{Gezari}, breakout shocks of Type II
SNe \cite{Schawinski,Soderberg:2008}, or mega-flares on otherwise normal,
main-sequence stars \cite{Djorgovski:2000}, etc.

We have been exploring variables and transients from the Palomar-Quest Sky
Survey \cite{Djorgovski:2008} (\url{http://www.palquest.org}) and the Catalina
Sky Survey (\url{http://www.lpl.arizona.edu/css}) in real-time and announcing
those via VOEventNet (\url{http://voeventnet.caltech.edu}).  Besides the
optical surveys (e.g. PTF, LSST, Pan-STARRS) there are many at other
wavelengths which will find transients and whose science can be enhanced by
real-time classification of these transients (e.g. Fermi, SKA, LOFAR, LISA to
name a few).

From targeted observing of small samples of a particular type of variable
objects or phenomena, the field has been moving towards a systematic
exploration of larger areas with a better time sampling and understanding of
finer details of these phenomena.  Rapid follow-up is essential for proper
understanding and scientific exploitation of the events varying on a short
time scale, or unusual classes of objects.  Many of these objects do not have
counterparts in archival image surveys like DSS, DPOSS, 2MASS etc. making the
discovery data points and any follow-up the sole data to go on.

An illustrative example was OT SNF143933+054631 discovered in the Point-and-Stare
data from Palomar-Quest by the LBNL SNF using archival comparison images.  The
initial SNIFs spectrum was highly unusual for a SN, and defied classification.
Using follow-up imaging with the Palomar 60-inch telescope and a spectrum at
Keck we were finally able to understand the nature of this peculiar SN
\cite{Mahabal:2006}.  Another case was SN2006lt, which turned out to be a rare SN of type Ib
\cite{Pecontal,Soderberg:2006}.  We may have discovered a class of SN associated
with faint dwarf galaxies in the process of looking for transients
\cite{Drake:2008a}.

SN and GRB are of course not the only transients for which unusual classes are
found.  CSS080924:233423+391423 seemed to be a simple flaring object until
follow-up imaging revealed that it had persisted 24 hours later
\cite{Mahabal:2008a} and Fig. 1).
A Palomar 200-inch spectrum revealed
numerous emission lines at zero redshift typical of Galactic dwarf novae.  The
atypically large variations at discovery remain unexplained \cite{Quimby}.
Yet another example is CSS080928:160837+041626, a possible high
amplitude ($\sim5 $ mags), long period variable, but with colors unlike one
\cite{Drake:2008b}.

The need for quick reporting and follow-up has resulted in
(1) the emergence of computer networks and protocols for collecting and
distributing streams of interesting events from large surveys -
the VOEventNet system
which serves events from a multitude of streams including 
Palomar-Quest survey and the Catalina Sky Survey
is a pertinent example \cite{Djorgovski:2008},
and (2) a number of robotic telescopes which can turn to a target very quickly
and provide crucial data for the classification of the events.

Here we describe the current status of the real-time event classification
effort. The endeavour is clearly applicable to other synoptic sky surveys. As
the event streams from synoptic sky surveys such as LSST and SKA increase,
real-time classification will become even more crucial as there will not be
enough facilities for follow-up observations making real-time classification a
key enabler of future synoptic astronomy.


A key difficulty of real-time classification of transients is the general lack
of available information initially available. A  transient detected by an
increase in brightness is often missing in archival sky surveys and may have
just a couple of relatively closely spaced observations in a couple of epochs
to go by.  Machine Learning methodologies including Support Vector Machines
(SVMs), Artificial Neural Networks (ANNs)  can be used as also Bayesian
classifiers including Naive Bayesian algorithms and Gaussian Process
regression \cite{Rausmussen}.

\begin{figure}
  \includegraphics[angle=0,height=.17\textheight]{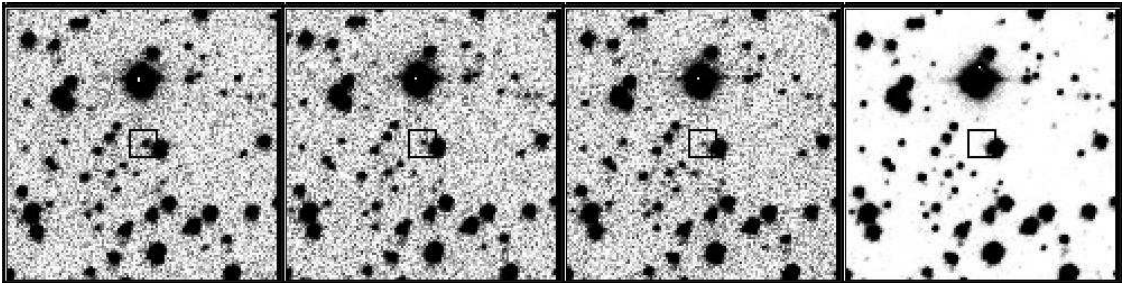}
  \caption{CSS080924:233423+391423 was classified as a transient based on its
flaring within the CSS images taken minutes apart. Not found in any archival
surveys except for the DSS N plate, it was still bright the next day. A
spectrum later revealed it to be a dwarf nova, but its rapid variation remains
unexplained. The three images on the left are individual CSS images taken on
Sep 24 2008 UT, and the one on the right is a co-added baseline image, also
from CSS.
	}
\end{figure}

\section{Methodology}
\subsection{Bayesian Event Classification}

Given a small number of observations, it is not generally possible to
unambiguously classify transients. The best approach in such a case is to
calculate the probabilities of the object belonging to different classes of
transients and then using objective criteria to determine if the probability
for the class of interest is high enough.
There will also be cases when none of the known classes is a good fit. This
type may perhaps turn out to be most interesting, with the transient being a
possible example of a new type of 
astronomical object or phenomenon.

We have described such a probabilistic method utilizing Naive Bayes method in
\cite{Mahabal:2008b}.
Briefly, the method involves building priors of different types of objects
with a large number of features. The object to be classified has a feature
vector which is decomposed into several independent blocks based on which
class is being considered. While this does not allow an exact membership to be
determined, obtaining approximate probabilities while circumventing the curse
of dimensionality as well as coping with several missing values makes this a
powerful method. See also \cite{HandYu,DomingosPazzani}.

\subsection{Machine Learning in Classification}

The machine learning, using ANNs/SVMs is more useful for dealing with
variables where more prior data is available \cite{Mahabal:2008b}. These too can deal with
incomplete information, at least partially, by training several sets of
quasi-independent classifiers and invoking the appropriate ones depending on
what information is available (magnitudes, colors, shape parameters etc.).

Another way we have started to use these techniques is to classify (and
eliminate) artifacts in real-time as data from telescopes are being processed.
Many artifacts, based on some features present in their signature and the fact
that they may not correspond to an object in the fiducial images, can get
initially flagged as transients. This technique helps remove them. More
details can be found in \cite{Donalek}. Similarly a very pertinent issue is
star-galaxy classification from multi-epoch data. Due to varying conditions
(seeing, airmass, extinction, filter) it is not possible to reproduce the same
class in each individual epoch. Combining the disparate data can be ideally
carried out using ANNs \cite{Donalek}.

\subsection{Feedback Incorporation}

Feedback incorporation \cite{Mahabal:2008b} is an important step since every
bit of additional information can potentially vindicate or contradict the
original classification (or at least revise probabilities). In addition, the
priors should also be updated so that classification in future has access to
the new information for the corresponding classes. This can be carried out
using either Expectation-Maximization algorithm \cite{Turmon} or kernel
density estimation \cite{Silverman}. The unknown parameters in these
approaches can be determined using known physical parameters for given classes
(e.g. SNs do not normally increase in brightness once they start fading, or
RR-Lyrae do not have their own host galaxies etc.) or additional observations
as they become available or labeling done by experts. This final bit plays an
important role in semantically connected portfolios built for each transient.
Such transient portfolios are structured yet flexible annotation mechanisms
including images and spectra besides comments. The collection of portfolios
can be indexed for ease of searching and execution of need-based services
\cite{Graham}.

\subsection{Follow-up Prioritization Engine}

Given the paucity of follow-up resources in the era of very large volume event
streams, matching transients to follow with such resources may well be the
most crucial step in advancing our knowledge about rare classes of quickly
fading objects. It will also be crucial in breaking the ambiguity between two
possible classes. An information-theoretic approach leading to the reduction
of final entropy is the best choice \cite{Loredo,Mahabal:2008b}.

\begin{figure}
  \includegraphics[angle=0,height=.4\textheight]{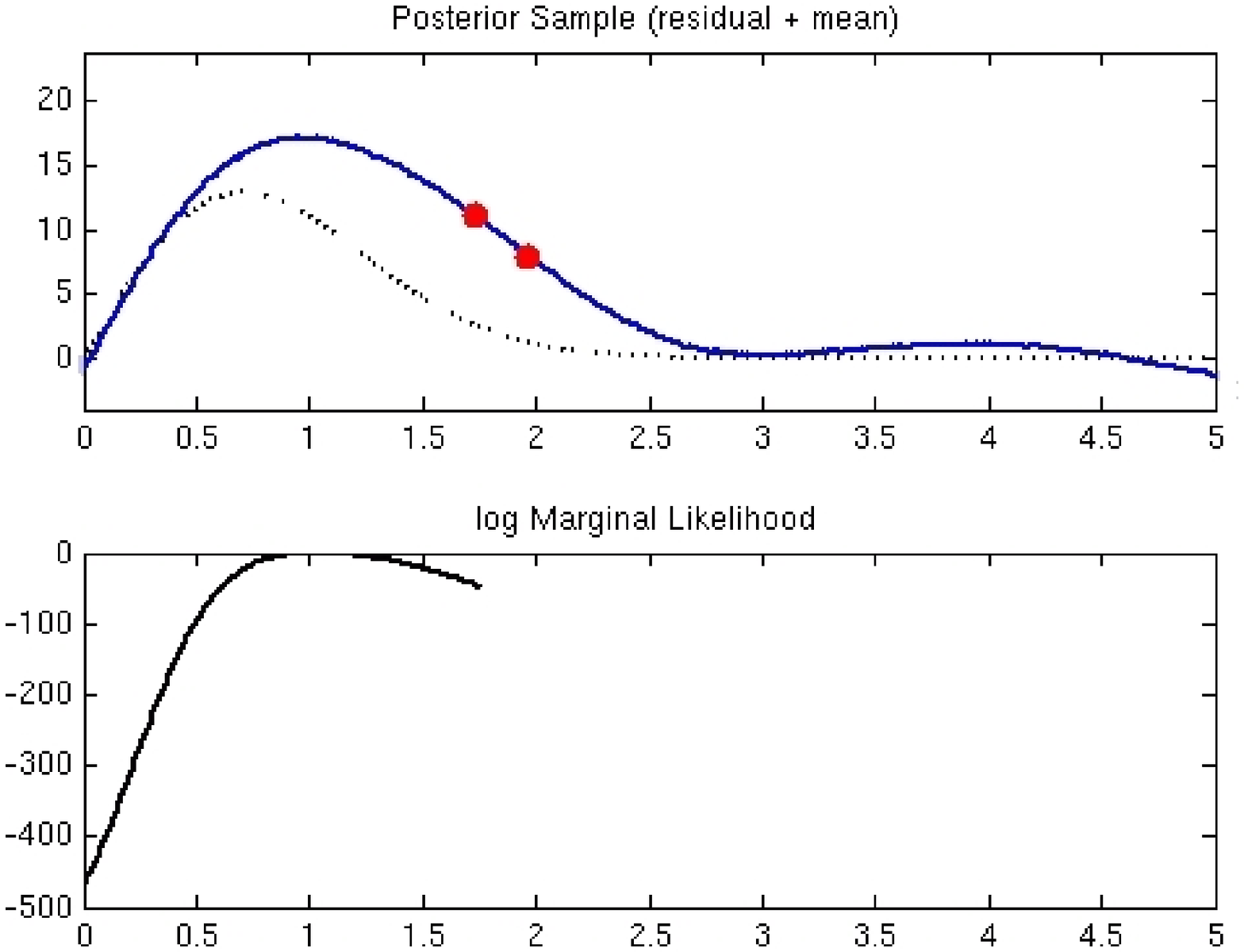}
\end{figure}
\begin{figure}
  \includegraphics[angle=0,height=.08\textheight]{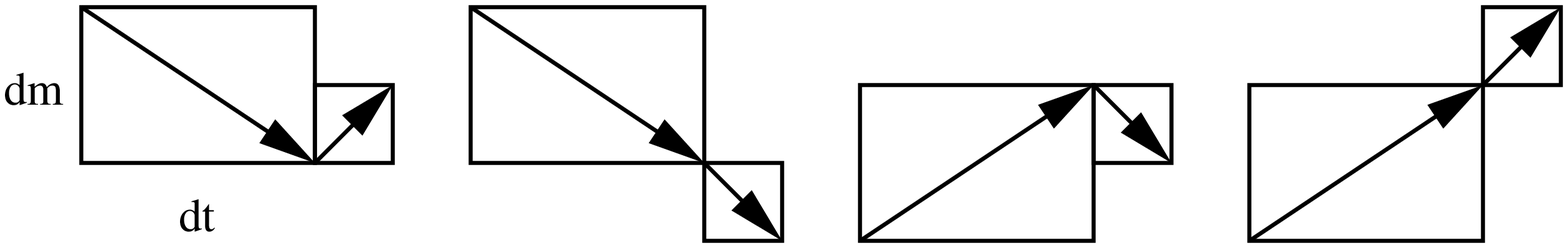}
  \caption{Gaussian Regression for a given $dm$ over the corresponding $dt$.
The second panel shows log marginal likelihood of the pair corresponding to
different parts of the given lightcurve. When more points are available for
comparison, it is easier to eliminate a larger number of previously competing hypotheses.
The boxes below show the distinct possibilities when three observations are
present. The lightcurve, within error-bars should accommodate the end-points of
the arrows to be a valid choice.
}
\end{figure}
\subsection{Gaussian Process Regression}

Given a small number of observations in a single band at different epochs, 
Gaussian Process Regression (GP) can be effective in determining class
membership. In this process priors are built from lightcurves of objects
belonging to a class of objects. For instance, consider just two epochs of a newly
discovered transient (Fig. 2). One can compare the two points against
different parts of a lightcurve in order to determine if there is a possible
fit and if so at what stage of evolution (e.g. periodicity). The lower panel
shows such a comparison with log marginal likelihood on the Y-axis. The more
observed points one has, the more stringent results one can obtain. Fig 3
shows different possibilities involving 3 points (i.e. $2\ dt$ and
corresponding $dm$). For a
given class of objects one can visualize a surface made of a grid of $dt$ and
$dm$
values which can then be used to determine class membership. However, owing to the
variation in objects belonging to a class, such a surface is somewhat fuzzy.

We have been conducting preliminary investigations using SN and Mira light
curves and will be extending it to other light curves. The results so far are
encouraging.

\subsection{VOEventNet}

VOEventNet  (\url{http://voeventnet.caltech.edu})
federates streams of astronomical events such that humans as well
as robotic telescopes can subscribe to the events. The events are available in
real-time in a standard format. The different streams allow subscribers to
choose events of their interest (e.g. SN, GRB, asteroids etc.). Google Sky
serves VOEventNet events under their "Current Sky Events" with links to
related technical and astronomical information. A color scheme allows more
recent events to stand out. AAVSO streams are also expected to be available
under VOEventNet soon.

The recently concluded Palomar-Quest survey and the ongoing Catalina Sky
Survey utilize VOEventNet to serve variables and transients in real-time. This
allowed quick follow-up in many cases and has resulted in datasets which would
otherwise be difficult to compile. For instance, during the first six months
of its real-time observations, CSS found 350 transients \cite{Drake:2008a}. 240 of these
were SN or CV and the remaining included AGNs, high proper motion stars,
highly variable stars as well as blazars and transients of an unknown nature.

We have been conducting follow-up of many of the interesting transients and
variables at the Palomar 60-inch telescope. These additional data aid in
classification as well as in the enhancement of priors for the Bayesian and
machine learning modules.

\begin{figure}
  \includegraphics[angle=0,height=.4\textheight]{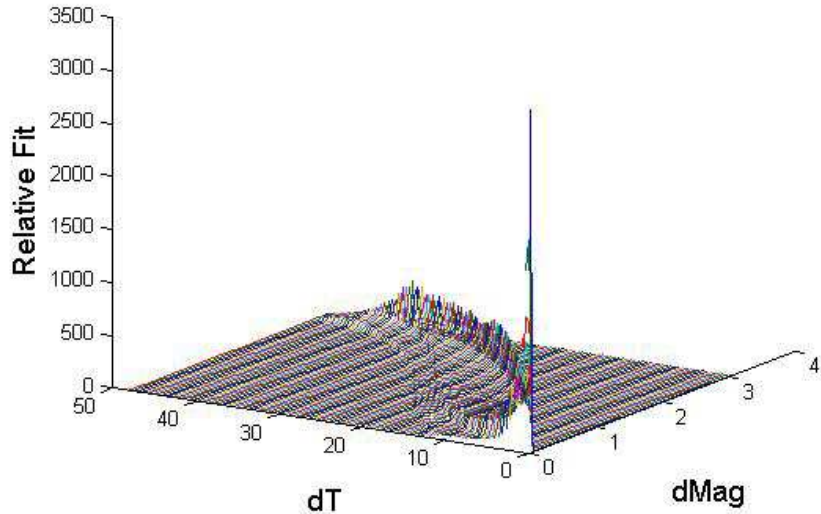}
  \caption{Given an object type one can build a surface indicating what likely
$dm$s at what $dt$s one is likely to encounter. One can then compare
any such pair against such a surface to see if the pair of points come from an
object of that type. In practice we of course do not have such a crisp
surface.}
\end{figure}

\section{Summary}
Many of the components described above are under development while some of
them exist
and prototypes have been successfully used to detect variables and transients in PQ and
CSS. But with event streams poised to become larger and larger, the methodology
and the architecture will have to be refined so that problem-specific,
on-demand computing services can compute/recompute virtual data according to a
particular algorithm as the need arises. This will have to be explicitly
linked using transient portfolios to various products like
"baseline sky" used in the detection and characterization of transient phenomena. 
Reaction and redirection of measurement processes are needed. 
A proper, scalable workflow incorporating these components will have
to be realized to make full use of the forthcoming event streams.

We have presented the status of real-time classification of events and
on-going developments including
Bayesian networks
and Machine Learning (ML) techniques.
Feedback from follow-up observations is necessary to improve priors
but will be increasingly scarce compared to the volume of event streams
making it important to continuously update training data sets.
The implementations of the classification methodology used in PQ/CSS along with
VOEventNet framework should help improve scientific returns from future
synoptic sky surveys.
\begin{theacknowledgments}
We are grateful to the staff of
Palomar Observatory for their help, and to our collaborators in PQ and CSS
survey teams.  This work was supported in part by the NSF grants
AST-0407448, AST-0326524, and CNS-0540369, and by the Ajax Foundation. 
A.K. and B.H. were supported in part by the Caltech SURF program.
\end{theacknowledgments}


\end{document}